\documentclass[conference]{IEEEtran}
\usepackage{cite}
\usepackage{amsmath,amssymb,amsfonts}
\usepackage{algorithmic}
\usepackage{graphicx}
\usepackage{textcomp}
\usepackage{xcolor}

\usepackage{subfiles}
 \usepackage{hyperref}
\usepackage{booktabs}
\usepackage{siunitx}
\usepackage{xcolor}
\usepackage{subcaption} 
\usepackage{enumitem}

\usepackage{booktabs, tabularx, makecell}
\usepackage[table]{xcolor}
\usepackage{colortbl}
\usepackage{multirow}
\usepackage{graphicx}
\usepackage{tikz}
\usepackage{multirow}
\usepackage{booktabs}
\usepackage{tcolorbox}
\usepackage{listings}
\definecolor{pairA}{HTML}{E8F1FD} 
\definecolor{pairB}{HTML}{E9F7EF} 
\definecolor{pairC}{HTML}{FFF4E5} 
\definecolor{pairD}{HTML}{F3E8FF} 
\definecolor{pairE}{HTML}{FFFDE7} 
\definecolor{pairF}{HTML}{FDEDEC}
\definecolor{posg}{RGB}{34,139,34} 
\definecolor{neug}{gray}{0.5}

\newcommand{\up}[1]{\textcolor{green!50!black}{\scriptsize$\uparrow$#1}}
\newcommand{\down}[1]{\textcolor{red!70!black}{\scriptsize$\downarrow$#1}}
\newcommand{\nochg}{\textcolor{gray}{\scriptsize--0}}
\newcommand*\circled[1]{\tikz[baseline=(char.base)]{
            \node[shape=circle,draw,inner sep=2pt] (char) {#1};}}

\definecolor{LightGray}{gray}{0.9}
\definecolor{Green}{rgb}{0,0.6,0}
\definecolor{LightPink}{rgb}{1.0,0.9,0.9}
\definecolor{Red}{rgb}{0.8,0,0}
\definecolor{Green}{rgb}{0,0.6,0}
\definecolor{codegray}{gray}{0.9}
\definecolor{keywordcolor}{rgb}{0.13,0.13,1.0}
\definecolor{stringcolor}{rgb}{0.58,0.0,0.82}
\definecolor{desiredcolor}{rgb}{1,0,0}  
\definecolor{assertcolor}{rgb}{0.8,0,0}
\definecolor{LightCyan}{rgb}{0.88,1,1}
\newcommand{\valdel}[2]{#1\;{\scriptsize #2}}

\makeatletter
\newcommand{\linebreakand}{%
  \end{@IEEEauthorhalign}
  \hfill\mbox{}\par
  \mbox{}\hfill\begin{@IEEEauthorhalign}
}
\makeatother

\newcommand{\tool}{\texttt{ConVerTest}}

\begin{document}

\title{Consistency Meets Verification: Enhancing Test Generation Quality in Large Language Models Without Ground-Truth Solutions}

\author{\IEEEauthorblockN{1\textsuperscript{st} Hamed Taherkhani}
\IEEEauthorblockA{\textit{York University} \\
Toronto, Canada \\
hamedth@yorku.ca}
\and
\IEEEauthorblockN{2\textsuperscript{nd} Alireza DaghighFarsoodeh}
\IEEEauthorblockA{\textit{York University} \\
Toronto, Canada \\
aliredaq@yorku.ca}
\and
\IEEEauthorblockN{3\textsuperscript{rd} Mohammad Chowdhury}
\IEEEauthorblockA{\textit{York University} \\
Toronto, Canada \\
ehsansh@yorku.ca}
\and
\IEEEauthorblockN{4\textsuperscript{th} Hung Viet Pham}
\IEEEauthorblockA{\textit{York University} \\
Toronto, Canada \\
hvpham@yorku.ca}
\linebreakand
\IEEEauthorblockN{5\textsuperscript{th} Hadi Hemmati}
\IEEEauthorblockA{\textit{York University} \\
Toronto, Canada \\
hemmati@yorku.ca}
}

\maketitle
\begin{abstract}

Large Language Models (LLMs) have significantly advanced automated test generation, yet existing methods often rely on ground-truth code for verification, risking bug propagation and limiting applicability in test-driven development. We present ConVerTest, a novel two-stage pipeline for synthesizing reliable tests without requiring prior code implementations. ConVerTest integrates three core strategies: (i) Self-Consistency (SC) to generate convergent test cases via majority voting; (ii) Chain-of-Verification (CoVe) for iterative, reasoning-guided code refinement; and (iii) a Dual Execution Agreement to cross-validate code and tests through consensus. Experiments on \textsc{BigCodeBench} and \textsc{Less Basic Python Problems (LBPP)} benchmarks demonstrate that ConVerTest improves test validity, line coverage, and mutation scores by up to 39\%, 28\%, and 18\% respectively over baselines. Our findings highlight ConVerTest as a robust solution for mitigating hallucinations and enhancing the reliability of autonomous software testing agents.

\end{abstract}

\section{Introduction}

LLMs have been successfully applied to a wide range of coding tasks in recent years, including automated test generation~\cite{10.1145/3660783,chen2024chatunitest,10.1109/TSE.2024.3382365,vikram2023can,plein2024automatic}. However, several challenges remain. A central difficulty in LLM-based test generation lies in verifying the correctness of the generated test cases~\cite{tian2025codehalu, liu2024exploring, chen2024chatunitest, 10329992}. For example, as shown in Table~\ref{table1}, only 35--51\% of tests generated by Gemma3.3 on BigCodeBench~\cite{zhuo2024bigcodebench} and 54--73\% of those generated by CodeQwen3 on LBPP~\cite{matton2024leakage} are valid. This raises a fundamental question: \textit{how to verify the verifier (the test cases)?}

Existing LLM-based test generation creates a "circularity of error" by using the implementation under test as the ground-truth oracle. This approach fails to detect faulty behavior because tests are generated to match current code logic rather than intended specifications. Some recent studies have raised hallucination issues that cause invalid test generation~\cite{10.1145/3661167.3661216, tian2025codehalu, liu2024exploring, chen2024chatunitest, 10329992, taherkhani2025automatedvalidationlanguagemodel, yuan2024evaluating,yang2024empirical}. Yang et al.~\cite{yang2024empirical} find that a substantial share of LLM-generated unit tests are invalid (often 34--62\%), mainly due to hallucination, where models invent code that looks plausible but is incorrect. They argue that post-processing strategies are needed to mitigate hallucinations.  

Other studies have explored generating valid LLM-based test cases, generally relying on execution of ground-truth code solutions for verifying tests~\cite{10.1145/3624032.3624035, li2023promptingcodeinterpreterwrite, li2024largelanguagemodelstest, 10820751}. However, generating tests from the code under test is risky because any bugs in the implementation can be reflected into the tests, making faulty behavior appear correct. Such tests capture what the code currently does rather than what it should do, which undermines their ability to detect errors. In some real-world scenarios like test-driven development, the code does not even exist yet, so relying on it for test generation is impossible. By contrast, generating tests from natural language description ties them to the intended behavior, ensuring tests remain independent, meaningful, and capable of revealing mistakes. Moreover, generating test cases from natural language is one of the core components of code generation agents, where they heavily rely on valid and useful test cases in their feedback loop to debug generated code solutions~\cite{shinn2024reflexion,zhou2023language,huang2023agentcoder,taherkhani2024epic,zhong2024ldb}. Generating tests without relying on code solutions is particularly challenging because there is no ground-truth implementation to provide feedback and refine the tests, so instead we need to develop a method that enforces consistency and mitigates hallucinations to synthesize valid and reliable test cases.

Hallucination mitigation techniques can be applied in multiple stages of LLM workflows, during and after generation~\cite{tonmoy2024comprehensivesurveyhallucinationmitigation}. In this paper, we propose {\tool}, which adapts and extends existing methods to improve the reliability and consistency of automated test generation in software development. Our method incorporates Self-Consistency (SC)~\cite{wang2022self} and Chain-of-Verification (CoVe)~\cite{dhuliawala2023chain} to mitigate hallucinations during the generation stage without relying on a ground-truth code solution. However, as shown in Section~\ref{Results}, enforcing consistency during generation alone is insufficient for producing valid tests. A post-generation step is therefore necessary to filter out remaining invalid tests. To this end, we adopt a dual-agreement strategy~\cite{chen2022codetcodegenerationgenerated} as a verification mechanism, whereby generated tests are cross-checked against candidate code solutions. Using this strategy, we systematically eliminate invalid tests, retaining only those that pass an additional verification. The verification step acts as a ground-truth proxy that eliminates invalid tests that could not be fixed in the generation stage and ensures a more rigorous and trustworthy test generation process. Finally, through ablation studies, we demonstrate the importance of each component in the pipeline. The quality of the test suites is evaluated at different stages to demonstrate the contribution of each step to the overall effectiveness of {\tool}.

Our empirical results demonstrate the following. First, SC improves test suite quality by increasing the proportion of valid tests by 7--19\% and significantly enhancing test adequacy metrics. Second, {\tool} further outperforms SC, improving test validity by 7--39\% through the elimination of invalid test cases during the post-generation verification stage. Third, ablation studies on {\tool}, conducted by progressively removing stages, reveal that all components contribute to classification accuracy and test suite quality. These findings confirm the complementary strengths of its core components: SC for convergent test generation, CoVe for iterative code refinement, and dual execution agreement for consensus-based verification.

\medskip
\noindent
The key contributions of this paper are as follows:
\begin{enumerate}[label=(\roman*)]
    \item We propose consistency-driven components inspired by the natural language processing domain and adapt them to synthesize reliable, high-quality code and test artifacts.
    \item We design and implement a novel two-stage pipeline that combines consistency-driven generation with a consensus-based verification stage, creating a more effective pipeline by drawing on hallucination mitigation strategies applied at both generation and post-generation stages.
    \item We conduct a large-scale empirical study on SOTA models and challenging benchmarks, demonstrating that {\tool} improves test validity rates by up to 39\% while also achieving high test adequacy. In addition, through comprehensive ablation studies, we quantify the individual contributions of each component and confirm that each part contributes to the performance of {\tool}.
\end{enumerate}

We also released the data and source code for our experiments to facilitate replication and extension by other researchers (\href{https://github.com/HamedTaherkhani/ConVerTest}{https://github.com/HamedTaherkhani/ConVerTest}).

\section{Background}
This study uses three methodologies: CoVe, SC, and Dual Execution Agreement.

\subsection*{On Chain-of-Verification}
CoVe~\cite{dhuliawala2023chain} is a deliberative, multi-stage reasoning method that improves factual accuracy and logical consistency in LLMs. The model (1) generates an initial draft, (2) creates verification questions to scrutinize its own claims and assumptions, (3) answers these questions (often independently to reduce bias) to detect and correct errors or ``hallucinations,'' and (4) synthesizes the results into a revised, verified response. Because functionally correct code generation is a known weakness of LLMs due to subtle logical flaws and hallucinations, we apply CoVe to code synthesis: draft \(\rightarrow\) interrogate \(\rightarrow\) refine. This structured self-correction shifts generation beyond pattern matching toward a process analogous to human developer review, yielding more robust and logically sound code.

\subsection*{On Self-Consistency}
SC~\cite{wang2022self} improves LLM reasoning by sampling multiple diverse reasoning paths for the same prompt and selecting the final answer via majority vote, based on the intuition that valid solution paths typically converge on a single correct outcome. This is well-suited to unit test generation because tests are convergent: they validate a single deterministic result for given inputs. We therefore generate a test stub, sample many completions, and choose the most frequent (most consistent) completion. This consensus-based selection helps filter stochastic generation errors, especially in assertion statements, and promotes convergence to the most probable assertions.

\subsection*{On Dual Execution Agreement}
Dual Execution Agreement selects the best code solution from multiple candidates by using LLM-generated test cases as a filter. Each candidate solution is executed against each generated test; solution--test pairs that pass are ``inliers.'' Solutions passing the same set of tests are grouped into consensus sets, which are scored by how many solutions and tests they contain. The final solution is chosen from the highest-scoring consensus set, motivated by the idea that correct solutions pass more tests and align with more similar solutions. Since CoVe and SC improve code and tests but do not guarantee correctness, we adapt Dual Execution Agreement as a post-generation quality gate to validate the generated tests themselves. The key principle is mutual corroboration: deliberatively produced code (CoVe) and consensus-derived tests are treated as independent high-confidence artifacts, and consensus-set behavior is used to distinguish valid from invalid tests.

\section{Approach}
\begin{figure*}[t]
  \centering
  \includegraphics[width=\textwidth]{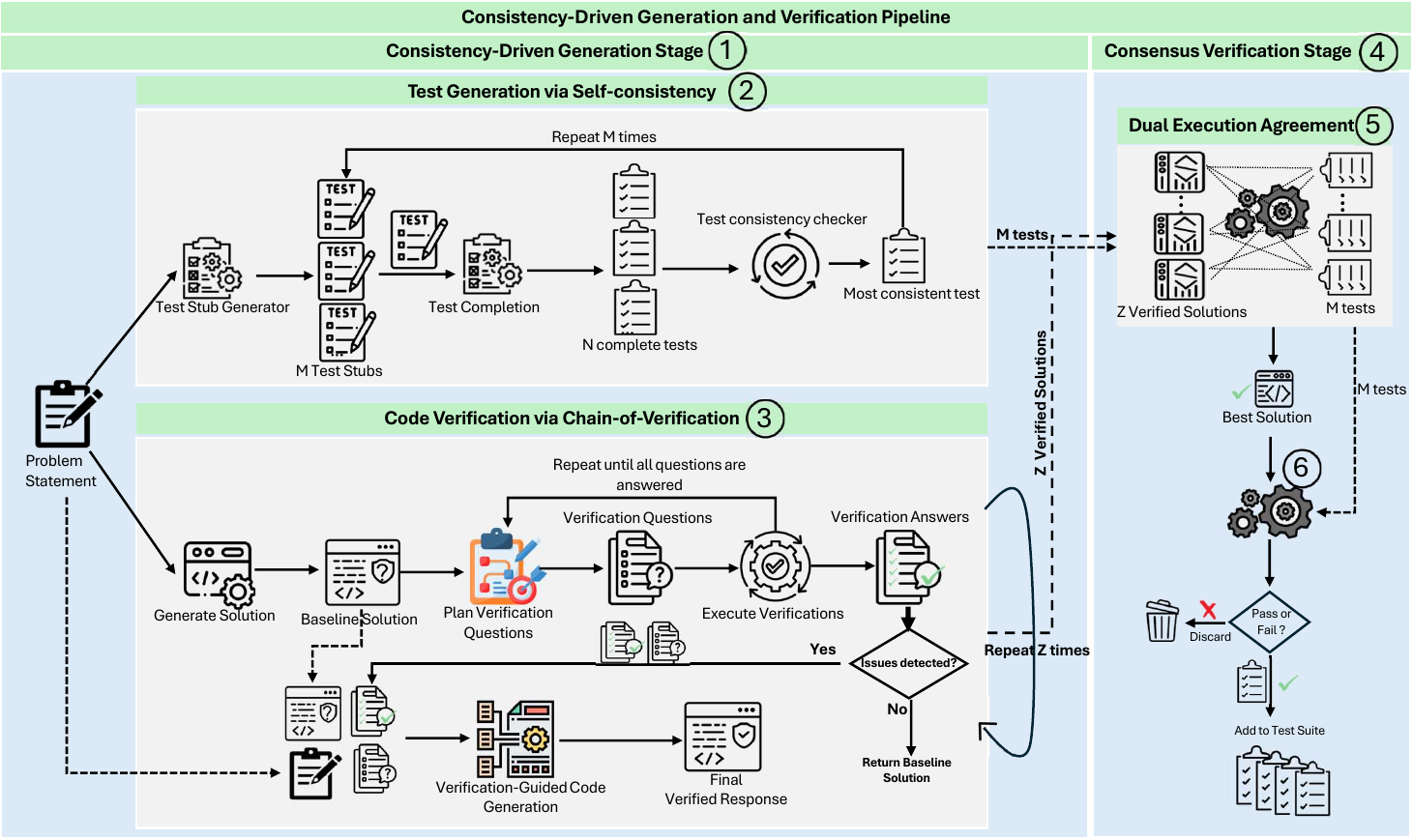}
  \caption{Overview of {\tool}. The pipeline operates in two stages: (i) the \textbf{Consistency-Driven Generation Stage}, which produces verified code candidates via \textbf{Chain-of-Verification} and generates diverse, convergent test cases through \textbf{SC}; and (ii) the \textbf{Consensus Verification Stage}, which applies a \textbf{Dual Execution Agreement} to cross-validate candidate solutions against generated tests and filter invalid ones.}
  \label{fig1}
\end{figure*}

{\tool} is illustrated in Figure~\ref{fig1}. {\tool} consists of two stages. 
The first stage, the \textit{Consistency-Driven Generation Stage} \circled{1}, is responsible for generating verified code candidate solutions as well as consistent test cases. 
The second stage, the \textit{Consensus Verification Stage} \circled{4}, is responsible for validating the generated tests using a consensus mechanism between the tests and the solutions. In the generation stage, {\tool} employs two simultaneous pipelines: one for test generation \circled{2} and another for code generation \circled{3}. The test generation pipeline produces $M$ test cases through SC, while the code generation pipeline produces $Z$ code candidate solutions using CoVe.  

The generated tests and code candidate solutions are then used in a dual-execution agreement process \circled{5} during the consensus verification stage to distinguish valid from invalid tests. The invalid tests are discarded, while the valid ones are incorporated into the test suite \circled{6}. 
Each stage of our pipeline is described in detail in the following subsections.  

\subsection{Consistency-Driven Generation Stage}
In Stage \circled{1}, two concurrent pipelines are employed: one responsible for generating consistent tests and the other for generating verified code. 
\subsubsection{Test Generation via SC} \circled{2}

In order to generate diverse and consistent test cases, we propose a two-stage SC approach. Generating a diverse set of test scenarios requires exploring a wide range of possibilities, including edge cases, boundary conditions, and typical usage patterns. In contrast, completing a given test stub with the correct assertion for a deterministic function is a simpler task. Attempting to perform both tasks in a single prompt can create a conflict; the LLM might default to simple, safe scenarios for which it can easily generate correct assertions, thereby sacrificing the diversity of the test suite. By decoupling these objectives, the pipeline enables the LLM to prioritize divergent, creative exploration during initial test stub generation. The process then transitions to a constrained, convergent reasoning phase in the second stage. This shift creates the ideal environment for SC, which leverages majority voting to identify a single, optimal completion from diverse samples.

\subsubsection*{Stage 1: Generation of Test Stubs}

At this stage, the language model must generate a set of $M$ diverse test stubs. A test stub is the skeleton structure of a unit test that defines the test scenario without concrete expected output values or specific assertion logic. It includes the test setup function, the input variable, and the function call. By directing the model toward this higher-level task, we guide it to produce a comprehensive set of test scenarios. This approach reduces the risk of omission of critical cases that may arise from computational complexity or hallucination in a single-shot generation.

The following is an example of a test stub:

\begin{lstlisting}[basicstyle=\footnotesize]
import unittest

class TestAnagramCombosMultiplePermutations(unittest.TestCase):
    def setUp(self):
        self.sub_words = ["ab", "cd", "ef"]
        self.whole_word = "abcdef"

    def test_multiple_permutations_possible(self):
        result = anagram_combos(self.sub_words, self.whole_word)
\end{lstlisting}
\subsubsection*{Stage 2: Completion and Refinement via SC}

Following the generation of the stubs, the second stage focuses on completing each stub using SC. For each of the $M$ stubs, the model is prompted $N$ times to generate a complete and concrete test function including specific expected outputs and assertions. This multiple-sampling strategy is the core of our SC implementation. It is motivated by the observation that while language models may produce inconsistent or erroneous outputs for complex tasks, they tend to converge on a correct or most probable answer when sampled repeatedly.

We select the most consistent completion for each stub using a consensus-based approach. The $N$ generated completions are first parsed to extract the underlying test logic and assertions. By analyzing the abstract syntax tree (AST) of each completion, we identify and group functionally identical test cases, abstracting away syntactic variations such as differing variable names or formatting. The test case with the highest frequency of occurrence among the $N$ samples is then chosen as the final test for that stub. This process is repeated for all $M$ stubs and produces a final suite of $M$ complete and consistent test cases.

This two-stage decomposition provides a significant advantage: it first ensures broad test coverage by generating distinct test scenarios and then refines each idea to produce a single, highly consistent and reliable completion of test scenario by implementing the expected output and assertion statements.

\subsubsection{Code Verification via CoVe} \circled{3}

To enhance the reliability and correctness of generated code, we employ a multi-stage, iterative refinement process termed \textbf{CoVe}. This methodology implements a systematic verification and correction loop. The core principle of CoVe is to model the rigorous process of human code review, where a solution is critically inspected against a set of targeted criteria, and feedback from this analysis directly informs subsequent revisions. The objective is to identify and rectify logical flaws, missed edge cases, and deviations from the problem description. 

A code solution can have multiple valid implementation pathways (e.g., an iterative vs. a recursive solution, or solutions using different library functions). A simple majority vote on the final code text would fail, as syntactically different but functionally identical solutions would not form a consensus. CoVe is better suited for this task because it focuses on iteratively refining a single reasoning path through self-interrogation and correction, mirroring a human code review process~\cite{10.1145/3664646.3664772}. This process improves the logical soundness and correctness of a solution regardless of its specific implementation details. The CoVe process is deconstructed into a sequence as illustrated in the following.

\subsubsection*{Baseline Solution Generation}
The process starts with the generation of an initial candidate solution, hereafter referred to as the \textit{baseline solution}. This solution is produced by an LLM in direct response to a given problem statement or prompt. This initial code serves as the starting point for the verification pipeline and is treated as a draft that requires verification rather than a final response.

\subsubsection*{Formulation of a Verification Plan}
Following the generation of the baseline solution, the next stage comprises creating a structured \textit{verification plan}. Rather than performing ad-hoc testing, this step programmatically generates a series of specific, analytical questions designed to probe the correctness of the baseline code. The generation of these questions is guided by the original problem statement (as the ground truth) and the baseline solution. The questions are designed to cover several critical aspects of the baseline solution:
\begin{itemize}
    \item \textbf{Correctness:} Assessing whether the code produces the correct output for a set of valid inputs.
    \item \textbf{Logic and Reasoning:} Examining the underlying algorithm for logical soundness and completeness.
    \item \textbf{Edge Case Handling:} Probing the solution's behavior with atypical or boundary inputs (e.g., empty lists, null values, zero).
    \item \textbf{Constraint Adherence:} Verifying that the solution respects all explicit and implicit constraints mentioned in the problem statement.
    \item \textbf{Robustness:} Evaluating the code's error-handling capabilities when presented with invalid or unexpected inputs.
\end{itemize}
This stage yields a list of targeted verification questions that constitute a formal checklist for subsequent analysis.

\subsubsection*{Execution of Verification Questions}
Each question from the verification plan is systematically addressed by analyzing the baseline solution's behavior. This analysis yields a corresponding set of \textit{verification answers}. These answers provide a detailed diagnosis of the code, explaining not only \textit{if} it passes a particular check but also \textit{why}, by reasoning about the expected versus actual behavior.

This stage results in a critical judgment of the solution's overall correctness and logic. Based on the collective evidence from the verification answers, a conclusion is made as to whether the code contains significant flaws that render it incorrect or incomplete.

\subsubsection*{Iterative Refinement via Verification-Guided Generation}
The outcome of the analysis stage dictates the next step in the pipeline. The process follows a conditional, iterative loop:
\begin{enumerate}
    \item If the analysis concludes that no issues are present, the verification loop terminates. The baseline solution is considered to have successfully passed the verification process and is returned as the final, validated output.
    \item If the analysis identifies one or more issues, the pipeline advances to a corrective step called \textit{verification-guided generation}. In this step, the LLM is prompted to produce a revised solution. The prompt is supplemented with the full context of the verification process: the original prompt, the flawed baseline solution, the set of verification questions, and the detailed verification answers that reveal the errors.
\end{enumerate}

This contextual information directs the model to produce a corrected solution that specifically addresses the identified shortcomings. 

The process is repeated $Z$ times, yielding $Z$ candidate solutions that are then used in the subsequent consensus-based verification step.

\subsection{Consensus Verification Stage}

The Consensus Verification Stage \circled{4} is the final phase of the pipeline and is responsible for selecting a single most reliable code solution from a set of previously verified candidates. It follows the principle of \emph{Dual Execution Agreement}, in which multiple candidate solutions and multiple generated test cases mutually validate one another to establish a strong consensus on correct program behavior. This stage takes two inputs: (i) a set of $Z$ candidate solutions that passed the initial checks in the Code Verification stage, and (ii) a set of $M$ generated test cases produced by the Test Generation stage. Each candidate solution is executed on every generated test case, producing an execution matrix of pass/fail outcomes for all $(\text{solution}, \text{test})$ pairs.

Using this matrix, the method forms \emph{agreement sets}, where each agreement set is a group of solutions that pass the exact same subset of test cases. For example, if Solution A and Solution B both pass Test 1 and Test 3 but fail all other tests, they are assigned to the same agreement set. This clustering identifies groups of solutions that exhibit identical behavior across the provided tests.

Each agreement set is assigned a consensus score to estimate reliability:
\[
\text{score} = (\text{number of tests passed}) \times \sqrt{\text{number of solutions in set}}.
\]
The first factor rewards sets whose solutions pass more tests, reflecting robustness and comprehensiveness. The second factor captures consensus strength: more solutions agreeing on the same behavior provides a stronger signal of correctness. The square root introduces diminishing returns so that very large groups of simple or mediocre solutions do not dominate smaller but higher-quality clusters. Agreement sets are ranked in descending order by this consensus score, and a representative solution from the top-ranked set is selected as the final solution. This ensures the selected output is both extensively validated by tests and supported by agreement among independently generated solutions. In addition to selecting the best solution, this stage evaluates the generated test cases themselves. The chosen best solution is treated as a reliable reference: each generated test is labeled \emph{valid} if it passes on this solution and \emph{invalid} if it fails. These labels enable computation of precision, recall, and F1 scores for the generated tests.

\section{Experiment Design}

\subsection{Models}
We employ three models in our experiments, selected to represent different providers (Alibaba, Google, and OpenAI), parameter scales, and abilities. Specifically, we include (i) a fine-tuned coding-oriented language model, (ii) a reasoning-focused model, and (iii) a general-purpose text generation model. The models are \textbf{qwen3-coder-480b-a35b-instruct\cite{qwen3technicalreport}, GPT-5-Mini-2025-08-07\footnote{\url{https://openai.com/index/introducing-gpt-5/}}, and Gemma-3.3-12b-it\cite{gemmateam2025gemma3technicalreport}}.




\subsection{Datasets}
We use two datasets in our evaluations: BigCodeBenchHard~\cite{zhuo2024bigcodebench} and Less Basic Python Problems (LBPP)~\cite{matton2024leakage}.

BigCodeBench is a large-scale benchmark for evaluating code generation LLMs through realistic programming challenges that require diverse function calls from 139 libraries across 7 domains and the ability to follow complex natural language instructions. BigCodeBench-Hard is an extended version with more difficult tasks involving higher complexity.

LBPP is a newly proposed benchmark for evaluating code generation in LLMs. It was created to address contamination issues in widely used benchmarks like HumanEval and MBPP. LBPP contains 161 manually crafted Python problems that are deliberately more difficult and designed to avoid overlap with existing datasets.

\subsection{Evaluation Metrics}
To assess the quality of test suites, we use two categories of metrics: (i) \textbf{Validity Rate (VR)} for measuring the proportion of valid test cases, and (ii) test adequacy metrics including \textbf{Mutation Score (MS)} and \textbf{Line Coverage (LC)}, both widely adopted in evaluating LLM-generated tests~\cite{shin2024assessing}.

\subsubsection{\textbf{Validity Rate (VR)}}
A test case is considered valid if it executes successfully against the ground-truth implementation without producing
any errors. Given functions \( F = \{ f_1, \dots, f_n \} \) and corresponding test sets \( T_i = \{ t_{i1}, \dots, t_{im_i} \} \), a binary function \( V(t_{ij}) \) indicates whether test case \( t_{ij} \) is valid for \( f_i \). The overall validity rate is defined as:
\[
\text{VR} = \frac{\sum_{i=1}^{n}\sum_{j=1}^{m_i} V(t_{ij})}{\sum_{i=1}^{n} m_i}.
\]

\subsubsection{\textbf{Mutation Score (MS)}}
Using the Mutmut tool, mutants are generated per function. If \( |M_i| \) is the number of mutants for \( f_i \) and \( K_i \) the killed mutants, the mutation score is:
\[
\text{MS} = \frac{\sum_{i=1}^{n} K_i}{\sum_{i=1}^{n} |M_i|}.
\]

\subsubsection{\textbf{Line Coverage (LC)}}
For each function \( f_i \), let \( L(f_i) \) denote its lines of code and \( L(t_{ij}) \) the lines covered by test \( t_{ij} \). The union of covered lines is:
\[
C(f_i) = \bigcup_{j=1}^{m_i} L(t_{ij}),
\]
with coverage ratio
\[
\text{Coverage}(f_i) = \frac{|C(f_i)|}{|L(f_i)|}.
\]
The average coverage is then:
\[
\text{LC} = \frac{1}{n}\sum_{i=1}^{n}\text{Coverage}(f_i).
\]

\subsubsection{\textbf{Precision and Recall}} 


In this paper, \textbf{precision measures the purity of the final test suite after {\tool} has filtered it}. It answers the question: ``Of all the test cases that {\tool} decided to \textit{keep}, what percentage are \textit{actually valid}?''. A high precision score means that the final test suite is clean and contains very few incorrect or invalid tests. \textbf{Recall measures how well {\tool} avoids throwing away valid tests}. It answers the question: ``Of all the test cases that were \textit{actually valid} to begin with, what percentage did {\tool} successfully \textit{identify and keep}?''. A high recall score means that the filtering process is not overly aggressive and preserves most of the useful, valid tests that were generated.




\subsection{Research Questions}

\begin{itemize}
    \item \textbf{RQ1: Effectiveness of SC.} To what extent does the two-stage Self-Consistency improve the validity and adequacy of synthesized tests compared to holistic and two-stage generation methods?
    \item \textbf{RQ2: Reliability of Consensus Verification.} How accurately can the consensus verification stage classify test cases as valid or invalid, and what is its impact on preserving high-quality tests within the final suite?
    \item \textbf{RQ3: Component Contribution.} What are the individual and collective impacts of CoVe, SC, and Two-Stage Generation (TSG) on the overall performance of {\tool}?
\end{itemize}

\section{Results}
\label{Results}
\begin{table}[t]
\centering
\footnotesize
\setlength{\tabcolsep}{2pt}
\renewcommand{\arraystretch}{0.5}
\caption{Holistic Test Generation (HTG), Two-Stage Test Generation (TSTG), and Self-Consistency Test Generation (SCTG) Comparison. All values are in percentages (\%)}
\label{table1}
\resizebox{\linewidth}{!}{
\begin{tabularx}{\linewidth}{l l l *{4}{>{\arraybackslash}X}}
\toprule
\textbf{Dataset} & \textbf{Approach} & \textbf{LLM} & \textbf{VR} & \textbf{LC} & \textbf{MS} & \textbf{\#Tests} \\
\midrule
\multirow{9}{*}{BCB} & HTG & \multirow{3}{*}{CodeQwen3} & 62  & 81  & 57  & 1808 \\
& TSTG & & 70 \up{8} & 85 \up{4} & 63 \up{6} & 2187 \\
& SCTG & & 76 \up{14} & 88 \up{7} & 67 \up{10} & 2187 \\
\cmidrule(lr){2-7}
& HTG & \multirow{3}{*}{Gemma3.3} & 35  & 54  & 64  & 1501 \\
& TSTG & & 45\up{10}  & 78\up{24}  & 71\up{7}  & 2097 \\
& SCTG & & 51\up{16}  & 82\up{28}  & 74\up{10}  & 2097 \\
\cmidrule(lr){2-7}
& HTG & \multirow{3}{*}{Gpt-5-Mini} & 61  & 72  & 50  & 3090 \\
& TSTG & & 68 \up{7} & 86 \up{14} & 62 \up{12} & 3109 \\
& SCTG & & 72 \up{11} & 89 \up{17} & 63 \up{13} & 3109 \\
\midrule
\multirow{9}{*}{LBPP} & HTG & \multirow{3}{*}{CodeQwen3} & 54  & 85  & 70  & 2115 \\
& TSTG & & 60 \up{6} & 91 \up{6} & 81 \up{11} & 2788 \\
& SCTG & & 73 \up{19} & 94 \up{9} & 85 \up{15} & 2788 \\
\cmidrule(lr){2-7}
& HTG & \multirow{3}{*}{Gemma3.3} & 38  & 60  & 44  & 1595 \\
& TSTG & & 45 \up{7} & 74 \up{14} & 51 \up{7} & 2938 \\
& SCTG & & 53 \up{15} & 76 \up{16} & 57 \up{13} & 2938 \\
\cmidrule(lr){2-7}
& HTG & \multirow{3}{*}{Gpt-5-Mini} & 70  & 71  & 49  & 2818 \\
& TSTG & & 74 \up{4} & 90 \up{19} & 65 \up{16} & 3019 \\
& SCTG & & 77 \up{7} & 91 \up{20} & 67 \up{18} & 3019 \\
\midrule
\end{tabularx}}
\end{table}

\subsection{ To what extent does the two-stage Self-Consistency improve the validity and adequacy of synthesized tests compared to holistic and two-stage generation methods?}
\label{RESRQ1}
In this RQ, we examine the strength of SC Test Generation (SCTG). To this end, we compare SCTG against two simpler baselines: Holistic Test Generation (HTG) and Two-Stage Test Generation (TSTG). In HTG, the entire test, including the test stub and the test output and assertions, are produced in a single step rather than generating the stub and the completion separately. In TSTG, we follow the two-stage procedure but, we sample only once \((N{=}1)\). This corresponds to a two-stage chain-of-thought approach.

The results in Table~\ref{table1} show that TSTG consistently produces more valid tests and higher-quality test cases in terms of line coverage (LC) and mutation score (MS) relative to HTG by improving validity by 4--10\%, LC by 4--24\%, and MS by 6--16\%. Adding SC to the process yields further gains: relative to HTG, SC improves validity by 7--19\%, and relative to TSTG, by 3--13\%. In addition, SC consistently increases the validity of synthesized tests and enhances their quality, improving LC by 7--28\% and MS by 10--18\%.

When comparing the performance of the different large language models, we observe distinct capability levels as shown in Table~\ref{table1}. CodeQwen3 shows particularly strong results on the BCB dataset, achieving the highest scores in all metrics with the SC approach (VR 76, LC 88, MS 67). Gpt-5-Mini demonstrates competitive performance, especially on the LBPP dataset where it achieves a high validity rate (VR 77). Gpt-5-Mini benefits the most from SC, showing the largest absolute gains in line coverage on both datasets when moving from the Holistic method to SC.

Gemma3.3 consistently scores lower than the other two models in both datasets. The performance gap is most evident on the more complex LBPP dataset. Despite its lower baseline performance, it is important to note that Gemma3.3 still shows significant improvement when using the TSTG and SCTG techniques. For example, its validity rate on LBPP improves from 38 to 53 with SCTG. This finding suggests that, while the choice of model is crucial for overall performance, employing advanced generation strategies provides substantial benefits for all models, regardless of their capability.
\begin{tcolorbox}
\textbf{Answer to RQ1:} SCTG is a highly effective technique for synthesizing higher-quality tests, significantly outperforming simpler baselines like HTG and TSTG. SC consistently improves test validity rate by up to 19\%, line coverage by up to 28\%, and mutation score by up to 18\%.
\end{tcolorbox}

\begin{table*}[t]
\centering
\footnotesize
\caption{Effectiveness of {\tool} in verifying test cases in terms of precision, recall, and F1 scores and retaining valuable tests in terms of LC and MS scores. Colored arrows show the percentage change from the \emph{Self-Consistency} to \emph{ConVerTest}. All values are in percentages (\%). A dash (-) indicates that the cell has no value.}
\label{table2}
\renewcommand{\arraystretch}{1}
\setlength{\tabcolsep}{3pt}
\begin{tabular}{l l l l l l l l}
\toprule
\textbf{Dataset} & \textbf{Approach} & \textbf{LLM} & \textbf{recall} & \textbf{precision (VR)} & \textbf{F1} & \textbf{LC} & \textbf{MS}\\
\midrule
\multirow{6}{*}{BCB} & Self-Consistency Tests & \multirow{2}{*}{Gpt-5-Mini} & -- & 72 & -- & 89 & 63 \\
 & ConVerTest & 
  & \valdel{95}{}
  & \valdel{86}{\up{16}}
  & \valdel{90}{}
  & \valdel{87}{\down{2}}
  & \valdel{61}{\down{2}} \\
\cmidrule(lr){2-8}
 & Self-Consistency Tests  & \multirow{2}{*}{CodeQwen3} & -- & 76 & -- & 88 & 67 \\
 & ConVerTest & 
  & \valdel{91}{}
  & \valdel{91}{\up{15}}
  & \valdel{91}{}
  & \valdel{87}{\down{1}}
  & \valdel{67}{\nochg} \\
\cmidrule(lr){2-8}
 & Self-Consistency Tests  & \multirow{2}{*}{Gemma3.3} & -- & 51 & -- & 82 & 74 \\
 & ConVerTest & 
  & \valdel{90}{}
  & \valdel{90}{\up{39}}
  & \valdel{90}{}
  & \valdel{81}{\down{1}}
  & \valdel{72}{\down{2}} \\
\midrule
\multirow{6}{*}{LBPP} & Self-Consistency Tests  & \multirow{2}{*}{CodeQwen3} & -- & 73 & -- & 94 & 85  \\
 & ConVerTest & 
  & \valdel{93}{}
  & \valdel{88}{\up{15}}
  & \valdel{90}{}
  & \valdel{93}{\down{1}}
  & \valdel{85}{\nochg} \\
\cmidrule(lr){2-8}
 & Self-Consistency Tests & \multirow{2}{*}{Gpt-5-Mini} & -- & 77 & -- & 91 & 67 \\
 & ConVerTest & 
  & \valdel{96}{}
  & \valdel{84}{\up{7}}
  & \valdel{89}{}
  & \valdel{91}{\nochg}
  & \valdel{65}{\down{2}} \\
\cmidrule(lr){2-8}
 & Self-Consistency Tests  & \multirow{2}{*}{Gemma3.3} & -- & 53 & -- & 76 & 57  \\
 & ConVerTest & 
  & \valdel{89}{}
  & \valdel{84}{\up{39}}
  & \valdel{86}{}
  & \valdel{74}{\down{2}}
  & \valdel{55}{\down{2}} \\
\bottomrule
\end{tabular}
\end{table*}
\subsection{How accurately can the consensus verification stage classify test cases as valid or invalid, and what is its impact on preserving high-quality tests within the final suite?}

To evaluate the effectiveness of the consensus verification stage in distinguishing between valid and invalid test cases, we use precision, recall, and F1 scores. Additionally, we assess the extent to which {\tool} retains high-quality tests by comparing the mutation score (MS) and line coverage (LC) before and after the verification phase.

\noindent\textbf{Precision as Validity Rate.} The precision metric is equivalent to the Validity Rate (VR) of the resulting test suite. That is, precision measures the proportion of tests that are valid among those retained after filtering. As shown in Table~\ref{table2}, {\tool} substantially improves VR across all models and datasets. For instance, on BigCodeBench, {\tool} boosts the VR from 72 to 86 for GPT-5-Mini and from 76 to 91 for CodeQwen3, representing absolute gains of 14\% and 15\% respectively. On the LBPP dataset, {\tool} improves the VR from 73 to 88 for CodeQwen3, and from 53 to 84 for Gemma3.3.

\noindent\textbf{Precision, Recall, and F1 Scores.} As shown in Table~\ref{table2}, {\tool} achieves consistently high precision and recall. For example, on BigCodeBench, {\tool} yields a precision of 91 and recall of 91 for CodeQwen3, resulting in an F1 score of 91. Similarly, for GPT-5-Mini, {\tool} attains a precision of 86 and recall of 95, confirming its ability to accurately identify valid test cases. On the LBPP dataset, {\tool} maintains high verification accuracy, achieving an F1 score of 90 for CodeQwen3 and 89 for GPT-5-Mini. Notably, the Gemma3.3 model, despite having lower baseline test quality, also benefits significantly from {\tool}, achieving an F1 score of 86 with a large precision gain (+39\%).

\noindent\textbf{Mutation Score and Line Coverage.}
While {\tool}'s verification stage is designed to remove invalid tests, it may also inadvertently discard a small number of valid tests. As a result, a slight drop in mutation score and line coverage is expected. However, the gain in VR outweighs this drop. This is evident in the results: for CodeQwen3 on BigCodeBench, the mutation score remains unchanged (67), while line coverage decreases slightly from 88 to 87. For GPT-5-Mini, mutation score drops from 63 to 61 and line coverage from 89 to 87. On the LBPP dataset, we observe similar trends: CodeQwen3 retains its mutation score (85), with only a 1\% drop in LC. For Gemma3.3, although the absolute scores are lower, the verification stage results in only a 2\% decrease in both metrics.

\noindent\textbf{Coverage-Validity Trade-off.} 
While Table~\ref{table2} indicates a marginal decrease in LC and MS after verification, this represents a shift from a \textit{high-quantity, low-reliability} test suite to a \textit{high-precision, high-reliability} one. The initial test suites often contain invalid tests that artificially inflate coverage metrics by executing code lines without actually verifying correct logic. By filtering these out, {\tool} ensures that the remaining LC and MS are derived solely from valid assertions. As shown in our results, the significant gains in Validity Rate (up to 39\%) far outweigh the negligible, expected reductions in coverage (1--2\%).

\begin{tcolorbox}
    \textbf{Answer to RQ2:} The consensus verification stage improves precision by up to 39\%. At the same time, it maintains high recall to avoid discarding valid ones, resulting in high F1 scores across all models. This comes at the cost of a minimal, expected drop in mutation score and line coverage, demonstrating that {\tool} successfully retains the vast majority of valuable tests while significantly enhancing the validity of the test suite.
\end{tcolorbox}

\begin{table*}[t]
\centering
\footnotesize
\setlength{\tabcolsep}{1pt}
\renewcommand{\arraystretch}{0.5}
\caption{Ablation study of {\tool}, demonstrating the contribution of each component. The configurations show the impact of progressively removing CoVe for code generation and SC and Two-Stage Generation (TSG) for test generation. Performance is measured across classification metrics (Recall, Precision/VR, F1-score) and test adequacy (Line Coverage and Mutation Score). All values are in percentages (\%)}
\label{table3}
\resizebox{\linewidth}{!}{
\begin{tabularx}{\linewidth}{c l l *{5}{>{\arraybackslash}X}}
\toprule
\textbf{Dataset} & \textbf{Approach} & \textbf{LLM} & \textbf{Recall} & \textbf{Precision (VR)} & \textbf{F1} & \textbf{LC} & \textbf{MS} \\

\midrule
\multirow{12}{*}[-5.5ex]{BCB} & ConVerTest & \multirow{4}{*}{CodeQwen3} & 91  & 91 & 91 & 87 & 67  \\
& ConVerTest w/o CoVe & & 85 \down{6}  & 91 \nochg & 88 \down{3} & 82 \down{5} & 61 \down{6}\\
& ConVerTest w/o CoVe \& SC & & 85 \nochg  & 88 \down{3} & 86 \down{2} & 81 \down{1} & 61 \nochg \\
& ConVerTest w/o CoVe \& SC \& TSG & & 85 \nochg  & 86 \down{2}  & 85 \down{1}  & 76 \down{5}  & 54 \down{7}  \\
\cmidrule{2-8}
\ & ConVerTest & \multirow{4}{*}{Gemma3.3} & 90  & 90  & 90  & 81  & 72  \\
& ConVerTest w/o CoVe & & 88 \down{2} & 90 \nochg & 89 \down{1} & 77 \down{4} & 68 \down{4}  \\
& ConVerTest w/o CoVe \& SC & & 88 \nochg & 88 \down{2} & 88 \down{1} & 77 \nochg & 68 \nochg  \\
& ConVerTest w/o CoVe \& SC \& TSG & & 88 \nochg & 85 \down{3} & 86 \down{2} & 51 \down{26} & 60 \down{8}  \\
\cmidrule{2-8}
& ConVerTest & \multirow{4}{*}{Gpt-5-Mini} & 95 & 86  & 90  & 87 & 61  \\
& ConVerTest w/o CoVe & & 93 \down{2} & 83 \down{3} & 88 \down{2} & 84 \down{3} & 58 \down{3} \\
& ConVerTest w/o CoVe \& SC & & 93 \nochg & 80 \down{3} & 86 \down{2} & 84 \nochg & 58 \nochg  \\
& ConVerTest w/o CoVe \& SC \& TSG & & 93 \nochg & 76 \down{4} & 84 \down{2} & 68 \down{16} & 47 \down{11}  \\
\midrule
\multirow{12}{*}[-5.5ex]{LBPP} & ConVerTest & \multirow{4}{*}{CodeQwen3} & 93 & 88 & 90  & 93 & 85 \\
& ConVerTest w/o CoVe & & 81 \down{12} & 89 \up{1} & 85 \down{5} & 86 \down{7} & 79 \down{6} \\
& ConVerTest w/o CoVe \& SC & & 77 \down{4} & 86 \down{3} & 81 \down{4} & 82 \down{4} & 75 \down{4} \\
& ConVerTest w/o CoVe \& SC \& TSG & & 81 \up{4}  & 89 \up{3}  & 85 \up{4}  & 78 \down{4}  & 69 \down{6}  \\
\cmidrule{2-8}
 & ConVerTest & \multirow{4}{*}{Gemma3.3} & 89 & 84 & 86 & 74 & 55 \\
& ConVerTest w/o CoVe & & 85 \down{4} & 84 \nochg & 84 \down{2} & 71 \down{3} & 49 \down{6} \\
& ConVerTest w/o CoVe \& SC & & 83 \down{2} & 84 \nochg  & 83 \down{1}  & 69 \down{2}  & 45 \down{4} \\
& ConVerTest w/o CoVe \& SC \& TSG & & 84 \up{1}  & 76 \down{8}  & 80 \down{3}  & 53 \down{16}  & 38 \down{7}  \\
\cmidrule{2-8}
 & ConVerTest & \multirow{4}{*}{Gpt-5-Mini} & 96 & 84 & 89 & 91 & 65 \\
& ConVerTest w/o CoVe & & 93 \down{3} & 84 \nochg & 88 \down{1}  & 89 \down{2} & 62 \down{3} \\
& ConVerTest w/o CoVe \& SC & & 92 \down{1}  & 82 \down{2} & 87 \down{1}  & 87 \down{2} & 60 \down{2} \\
& ConVerTest w/o CoVe \& SC \& TSG & & 92 \nochg  & 80 \down{2}  & 86 \down{1}  & 65 \down{22}  & 42 \down{18}  \\
\bottomrule
\end{tabularx}}
\end{table*}

\subsection{What are the individual and collective impacts of CoVe, SC, and Two-Stage Generation (TSG) on the overall performance of {\tool}?}

To investigate the contribution of each component in {\tool}, we performed an ablation study in which we progressively removed or altered the test and code generation components, while observing their effects on test suite metrics and on the classification ability of {\tool} (recall, precision, and F1 scores). Table~\ref{table3} reports the comparative results across different configurations on both BigCodeBench (BCB) and LBPP datasets using multiple LLMs.

It is important to note that the precision metric is equivalent to the Validity Rate (VR) of the resulting test suite. The precision also represents the precision of the dual-agreement classification ability in {\tool}. Thus, they are essentially the same in this table.

\noindent\textbf{{\tool} without CoVe.} In the first ablation study, the only change relative to the previous configuration is the removal of CoVe. Instead, we use vanilla code generation. In vanilla code generation, we generate the baseline solution, as we did with CoVe, but treat it as the final solution without applying planning or verification. This leads to a significant drop in the recall of dual agreement verification because the candidate solutions generated by vanilla code generation are much less accurate than those produced with CoVe. As a result, {\tool} incorrectly classifies more valid tests as invalid and removes them from the test suite.

Discarding more valid tests sharply decreases both the mutation score (MS) and line coverage (LC) of the resulting test suites. Specifically, LC decreases by 2--7\% and MS by 3--6\%. Thus, replacing CoVe with vanilla code generation degrades the quality of the generated code and reduces the recall of {\tool}'s classification ability. This decrease is consistent across all dataset LLM combinations.

\noindent\textbf{{\tool} without CoVe and SC.} In the second ablation study, we remove only the SC test generation component from the previous configuration. In this setting, CoVe is not used for test generation, and SC is excluded as well. However, we retain the Two-Stage test generation (TSG) and replace SC with CoT prompting for test generation. In 5 out of 6 cases, this reduces both the precision of the approach and the validity rate of the test suites compared to the previous configuration. In addition, both line coverage (LC) and mutation score (MS) decrease in most cases. This is expected because, as shown in Table~\ref{table1}, SC test generation yields higher-quality and more adequate tests than TSG. The decline in the precision of {\tool} under this configuration shows that test case quality directly affects {\tool}'s ability to distinguish invalid tests, and improving test quality enhances {\tool}'s effectiveness in verifying test cases.

This configuration consistently reduces both validity and adequacy compared to the previous setup. For example, on BCB with CodeQwen3, precision drops from 91 to 88, while LC falls from 82 to 81. On LBPP, CodeQwen3 shows declines in both recall and precision, as well as decreases in MS and LC (from 86 to 82 and from 79 to 75, respectively). Gemma3.3 exhibits a slightly different pattern: precision remains the same, but recall drops from 85 to 83, and LC and MS decrease from 71 to 69 and from 49 to 45, respectively. These results demonstrate that replacing SC with Two-Stage test generation (TSG) substantially degrades adequacy metrics, reduces the validity rate of the test suites, and weakens {\tool}'s predictive ability compared to configurations that incorporate SC in the test generation process.

\noindent\textbf{{\tool} without CoVe and SC and TSG.} The final ablation study evaluates {\tool} without CoVe, SC, and TSG. In this setting, we rely solely on holistic test generation and vanilla code generation, meaning that no verification or consistency techniques are applied during the code or test generation process.

Removing CoVe and SC results in a non-trivial decrease in test adequacy metrics, specifically line coverage (LC) and mutation score (MS). LC decreases by 4--26\%, while MS decreases by 6--18\%. For example, on LBPP–Gpt-5-Mini, LC drops by 22 (absolute) and MS by 18 (absolute). This is expected, as the quality of holistic test generation is considerably lower than that of TSG or SC. The reduced quality of holistic test generation also degrades the classification ability of {\tool}. In 5 out of 6 cases, the precision of {\tool} declines. For instance, on Gemma3-LBPP, LC decreases by 16, MS by 7, and precision by 8.

\begin{tcolorbox}
\textbf{Answer to RQ3:} Each component of {\tool} has a significant impact on its overall performance. The ablation study reveals that CoVe is critical for producing reliable code solutions for the verification stage, and its removal substantially degrades recall and test suite adequacy. SC plays a crucial role in enhancing the initial quality and validity of generated tests, as its absence leads to lower precision and reduced adequacy scores.
\end{tcolorbox}

\section{Discussion}

While our results demonstrate that \texttt{ConVerTest} can effectively filter invalid test cases, it is crucial to understand the underlying reasons why LLMs generate such tests in the first place. To this end, we conducted a qualitative analysis to categorize the types of errors that lead to invalid test cases.

We randomly sampled $100$ invalid tests generated from both the BigCodeBench and LBPP datasets. Two graduate students with experience in software engineering independently categorized the root cause of failure for each test. A third graduate student then reviewed the classifications and resolved any conflicts, ensuring a reliable categorization. The percentage of agreement between the first two annotators is $79\%$. The results of this analysis are summarized in Table \ref{tab:invalid_test_categories}.

\begin{table}[h!]
\centering
\setlength{\tabcolsep}{2pt}
\renewcommand{\arraystretch}{1}
\caption{Categorization of Root Causes for Invalid LLM-Generated Test Cases}
\label{tab:invalid_test_categories}
\begin{tabular}{@{}lc@{}}
\toprule
\textbf{Category of Invalidity} & \textbf{Frequency (\%)} \\
\midrule
Expecting behavior not requested in the spec & 28 \\
Wrong API / function call & 19 \\
Misunderstanding the Function Logic from NL Description& 15 \\
Misread edge-case logic & 12 \\
Ambiguities or Incompleteness in the Function Docstring & 12 \\
Violating stated input constraints / preconditions & 7 \\
Output-format inconsistency & 7 \\
\bottomrule
\end{tabular}
\end{table}

Our analysis reveals several key insights. The most prevalent issue, accounting for 28\% of invalid tests, is the LLM generating tests that \textbf{expect behavior not requested in the specification}. This form of hallucination, where the model makes assumptions or invents requirements, is a significant challenge. For example, a model might assert that a function should sort a list in-place when the docstring only specifies that a sorted list should be returned.

The second most common failure (19\%) is the use of a \textbf{wrong API or function call}. This is particularly common in complex benchmarks like BigCodeBench, where models must correctly utilize functions from various libraries. These errors range from calling a non-existent function to providing incorrect arguments, indicating a weakness in the model's knowledge of specific library APIs.

A significant portion of the errors stems from a deeper semantic misunderstanding. Failures to correctly interpret \textbf{natural language logic} (15\%) and \textbf{edge-case logic} (12\%) show that LLMs can struggle with complex requirements. Furthermore, we found that in 12\% of the cases, the invalidity could be traced back to \textbf{ambiguities or incompleteness in the function docstring} itself. This suggests that the quality of the dataset is a critical factor and that LLMs are susceptible to generating incorrect tests when the function docstring is not perfectly clear.

The remaining categories, \textbf{violating input constraints} (7\%) and \textbf{output-format inconsistency} (7\%), represent more direct logical contradictions but are less frequent.

These findings strongly support the need for a robust verification mechanism like \texttt{ConVerTest}. The prevalence of semantic errors (28\% + 15\% + 12\%) underscores why simple syntactic checks are insufficient. The dual execution agreement at the core of \texttt{ConVerTest} is well-suited to detect these discrepancies. When a test expects a behavior not specified in the docstring, it will likely fail against a code solution generated via the rigorous CoVe process, which is designed to adhere closely to that same docstring. By establishing a consensus between independently generated code and tests, \texttt{ConVerTest} effectively creates a proxy for ground truth that can identify and discard tests suffering from these common forms of LLM hallucination and misinterpretation.

\section{Related Work}
\subsection{Test case generation}

In recent years, LLMs have opened new possibilities for automatically generating unit tests, making them easier for developers to create while also reducing development time. By learning from large collections of developer-written tests, these models can create test cases that go beyond what traditional techniques offer. Researchers have tried different ways to improve the quality of these tests, including pretraining and fine-tuning on code-related tasks \cite{tufano2020unit,alagarsamy2024a3test,shin2024domain,rao2023cat}, reinforcement learning \cite{steenhoek2025reinforcement}, better prompt design \cite{zhang2023well,dakhel2024effective,chen2024chatunitest}, the use of documentation~\cite{plein2024automatic,vikram2023can}, and even integration with search-based methods~\cite{lemieux2023codamosa}.

At the same time, challenges remain. Studies show that LLM-generated tests often struggle with correctness~\cite{li2023prompting,guilherme2023initial,yuan2023no} and may provide limited coverage on some benchmarks. This has led to direct comparisons with well-established tools such as EvoSuite~\cite{bhatia2024unit,10.1109/TSE.2024.3382365} and Pynguin~\cite{10.1109/ICSE48619.2023.00085,bhatia2024unit}, which underscores the need for continued refinement.

Several recent projects demonstrate how the field is moving forward. Liu et al.~\cite{liu2024llm} introduced AID, which combines LLMs with differential testing to reveal faults in programs that already pass conventional test suites. Alagarsamy et al.~\cite{alagarsamy2024a3test} developed A3Test, which applies domain adaptation to produce assertion-informed test cases with consistent naming. ChatTester~\cite{10.1145/3660783} focuses on refining the tests produced by ChatGPT through iterative feedback loops, improving their precision. In another study, Ouédraogo et al.~\cite{ouedraogo2024large} compared multiple LLMs with EvoSuite, examining their strengths in coverage, readability, and correctness.

Before these advances, traditional methods relied mainly on search-based~\cite{harman2009theoretical,delgado2022interevo}, constraint-based~\cite{xiao2013characteristic}, or random-based~\cite{pacheco2007feedback} strategies to maximize code coverage. Although effective for this narrow purpose, they often resulted in test cases that were difficult to maintain. The move toward LLM-driven approaches marks an important step forward, even if more work is needed to address current limitations.

\subsection{Inconsistency in LLMs}

Recent research has increasingly framed inconsistency as a fundamental limitation of LLMs, with multiple strategies proposed to improve robustness and reasoning reliability. Dhuliawala et al.~\cite{dhuliawala2023chain} introduce \textit{CoVe}, where models explicitly plan and answer fact-checking questions to reduce hallucinations. Similarly, He et al.~\cite{he2024retrievingrethinkingrevisingchainofverification} extend this idea in retrieval-augmented generation, proposing \textit{CoV-RAG}, which integrates verification to refine both external retrieval and internal generation. Complementing verification-based methods, Zhao et al.~\cite{zhao-etal-2024-improving} propose \textit{Consistency Alignment} as a two-stage training framework: instruction augmentation followed by consistency-based preference alignment, aiming to improve robustness to paraphrased prompts. 

Another prominent direction builds on \textit{SC}~\cite{wang2022self}, originally formulated as sampling multiple reasoning chains and aggregating by majority voting. Chen et al.~\cite{chen-etal-2024-self-para} propose \textit{Self-Para-Consistency}, leveraging paraphrased inputs to generate diverse but high-probability reasoning paths, thus reducing the heavy sampling cost of standard SC. Huang et al.~\cite{huang-etal-2024-mirror} extend this idea with \textit{Mirror-Consistency}, which incorporates minority and inconsistent answers as valuable signals to calibrate confidence and reduce overconfidence. Wang et al.~\cite{wang2024softselfconsistencyimproveslanguage} propose \textit{Soft Self-Consistency (Soft-SC)}, replacing hard majority voting with probability-weighted scoring, enabling more efficient selection in tasks with diverse valid outputs. Further, Prasad et al.~\cite{prasad2025selfconsistencypreferenceoptimization} embed SC into training through \textit{Self-Consistency Preference Optimization (SCPO)}, aligning models to prefer consistent responses even without gold supervision. These approaches converge on the principle that internal disagreement among model samples, rather than being discarded, can be leveraged as a structured signal for reliability.

\section{Limitations and Threats to Validity}

While {\tool} shows promising results in generating valid and reliable test cases, several limitations must be considered. We also highlight how our design choices mitigate or alleviate these threats.

\textbf{Internal Validity:}
Our approach relies on specific sampling strategies (e.g., number of test completions in SC and verification rounds in CoVe). Different hyperparameter choices may lead to different outcomes. Moreover, our consensus-based verification assumes that convergence across independently generated artifacts correlates with correctness, which may not hold if multiple incorrect yet internally consistent outputs are produced. To mitigate this, we conducted extensive ablation studies to show that each component (SC, CoVe, and Dual Execution Agreement) individually contributes to the overall results, and their combination consistently achieves the best results.

\textbf{External Validity:}
Our experiments focus on Python programming tasks from BigCodeBench and LBPP. Although diverse, these datasets may not fully represent all programming paradigms, strongly-typed languages, or large-scale system testing. Thus, generalizability to other languages or real-world projects remains for future work. We deliberately selected benchmarks that are widely used, challenging, and designed to avoid contamination (LBPP) or trivial solutions (BigCodeBench-Hard). This choice increases external validity compared to smaller or potentially leaked datasets.

\textbf{Model and Data Bias:}
The performance of {\tool} is constrained by the underlying LLMs, which vary in coding capabilities and may contain biases from their training data. For instance, weaker models (e.g., Gemma3.3) show lower baseline performance. Moreover, although LBPP was designed to mitigate training-data leakage, residual contamination cannot be entirely ruled out. To mitigate this, we explicitly tested with three heterogeneous models of different strengths (a code-specialized LLM, a reasoning-focused model, and a general-purpose model). The consistent improvements across all models demonstrate that {\tool} is not narrowly tuned to one configuration or LLM, and that its benefits generalize despite underlying model biases.

\textbf{Conclusion Validity and Practical Applicability:}
{\tool} has not yet been evaluated in continuous integration pipelines, test-driven development workflows, or at industrial scale. Its multi-stage pipeline introduces additional computational cost due to repeated sampling, verification, and dual execution, which may limit deployment in resource-constrained environments. We measured adequacy preservation (mutation score and line coverage) and showed that {\tool} discards only a small fraction of valid tests while substantially improving overall validity. This trade-off suggests that, despite higher computational cost, the resulting test suites are more trustworthy and therefore more cost-effective in the long run.

\section{Conclusion and Future Work}
In this paper, we introduced \textbf{{\tool}}, a novel approach that addresses the challenge of generating valid and reliable test cases with LLMs without relying on ground-truth implementations. By combining three complementary strategies---Self-Consistency for convergent test generation, Chain-of-Verification for iterative code refinement, and Dual Execution Agreement for consensus-based verification---{\tool} systematically mitigates hallucinations and enhances the quality of synthesized test suites. Our empirical evaluation on \textsc{BigCodeBench-Hard} and \textsc{LBPP} demonstrates that {\tool} achieves superior performance compared to holistic and two-stage baselines, consistently improving validity rates, line coverage, and mutation scores. Furthermore, ablation studies confirmed the critical contribution of each pipeline component, validating the necessity of integrating consistency-driven generation with consensus verification.

While {\tool} advances the reliability of automated test generation, several opportunities remain for future exploration. First, extending our approach to other programming languages and multi-modal tasks, such as tests derived from API specifications or natural language documentation, would broaden its applicability. Second, incorporating semantic and dynamic analysis techniques into the verification stage could further reduce false positives and capture deeper logical flaws that execution-based verification may miss. Third, adapting {\tool} for interactive and incremental development workflows, such as test-driven development or continuous integration pipelines, would enable tighter integration into real-world software engineering practices. Finally, future work should explore hybrid strategies that combine {\tool} with search-based or formal verification approaches.
\section{Data Availability}
We released the data and source code for our experiments to facilitate replication and extension by other researchers (\href{https://github.com/HamedTaherkhani/ConVerTest}{https://github.com/HamedTaherkhani/ConVerTest}).
\bibliographystyle{IEEEtran}
\bibliography{main}
\end{document}